\documentclass{PoS}

\usepackage{amsmath,amssymb}

\title{Applying Integrability to Gauge Theories}

\ShortTitle{Applying Integrability...}

\author{\speaker{Peter Orland}\thanks{Supported in part by NSF Award No. PHY-0855387,  and
a PSC-CUNY Research Award.}\\
        Baruch College and the Graduate School and University Center, the City University of New York, New York, NY\\
        E-mail: \email{orland@nbi.dk}}

\abstract{Lattice Yang-Mills theories in any dimension may be regarded as coupled $1+1$-dimensional integrable field theories. These integrable systems decouple at large center-of-mass energies, where the action becomes effectively anisotropic. This effective action is the high-energy center-of-mass limit of the gauge theory. In $2+1$ dimensions, the quark-antiquark potential and the mass spectrum can be calculated, using the exact $1+1$-dimensional S-matrix and form factors. The methods are quite similar to those applying integrability in statistical and condensed-matter physics. The high-energy anisotropic action at one loop in $3+1$ dimensions has been found using a Wilsonian renormalization method. We briefly discuss the isotropic theory in $2+1$ dimensions and
the connection with soft scattering in $3+1$ dimensions.}

\FullConference{The XXVIII International Symposium on Lattice Field Theory, Lattice2010\\
		June 14-19, 2010\\
		Villasimius, Italy}

\begin{document}

\section{Introduction}

The ideas discussed here were motivated by the following observation. Consider the simple two-coupling version of Yang-Mills theory, whose Minkowski-space Lagrangian is 
\begin{eqnarray}
{\mathcal L}=\frac{1}{2(g_{0}^{\prime})^{2}}{\rm Tr}\,\,F_{01}^{2}+
\frac{1}{2g_{0}^{2}}\, {\rm Tr}\,\,F_{02}^{2}- \frac{1}{2g_{0}^{2}}{\rm Tr}\,\,F_{12}^{2}, 
\label{cont-lag-2+1}
\end{eqnarray}
where $F_{\mu\nu}$ is the Yang-Mills field strength $F_{\mu\nu}=\partial_{\mu}A_{\nu}
-\partial_{\nu}A_{\mu}-i [A_{\mu}, A_{\nu}]$ and the gauge field $A_{\mu}$ is in the Lie algebra of 
SU($N$).  When regularized on a lattice, this theory can be shown to confine quarks and have a mass gap, when $g_{0}^{\prime}$ is sufficiently small, for any $g_{0}$ \cite{orland2+1}. 

Using simple arguments, the mass gap and quark-antiquark potential were found for 
$(g_{0}^{\prime})^{2} \ll \frac{1}{g_{0}}e^{-4\pi/(g_{0}^{2}N)}$. This was done by exploiting the connection between the lattice anisotropic gauge theory and
the ${\rm SU}(N)\times {\rm SU}(N)$-symmetric sigma model in $1+1$ dimensions. Subleading corrections were found to these physical quantities \cite{hor-string, mass-spectrum,vert-string}, using the exact S-matrix \cite{abda-wieg} and form factors \cite{kar-wiesz}. It is noteworthy that these are entirely weak-coupling
methods. The main technical problem is the presence of a dimensional cross-over between 
$1+1$-dimensional behavior and that of the isotropic gauge theory \cite{hor-string}.

In fact,
(\ref{cont-lag-2+1}) is the high-energy center-of-mass form of the $2+1$-dimensional gauge field 
theory Lagrangian. The high-energy limit can be obtained by a longitudinal rescaling $x^{0, 1}\rightarrow \lambda x^{0,1}$, 
$x^{2}\rightarrow \lambda x^{2}$, with $\lambda\ll1$ and $g_{0}^{\prime}=\lambda g_{0}$. In the more physical case of $3+1$-dimensional QCD, such a rescaling is 
$x^{0,3}\rightarrow \lambda x^{0,3}$, 
$x^{1,2}\rightarrow x^{1,2}$ and the effective Lagrangian is \cite{VV, MV}
\begin{eqnarray}
{\mathcal L}=\frac{1}{2g_{0}^{2}}\sum_{j=1,2}{\rm Tr}\,\,F_{0j}^{2}+\frac{1}{2(g_{0}^{\prime})^{2}}{\rm Tr}\,\,F_{03}^{2}
-\sum_{j=1,2}\frac{1}{2g_{0}^{2}}{\rm Tr}\,\,F_{j3}^{2}- \frac{1}{2(g_{0}^{\prime\prime})^{2}}{\rm Tr}\,\,F_{12}^{2}, \label{cont-lag-3+1}
\end{eqnarray}
where $g_{0}^{\prime\prime}=\lambda^{-1}g_{0}$. The structure of hadrons in a lattice version of this effective theory
was discussed in Reference \cite{orland-3+1}. The actual high-energy effective action must include anomalous powers of $\lambda$ in the coefficients of the field strength, which have been found to one loop \cite{orland-xiao}. In the limit of small $\lambda$, this theory can also be shown to confine
quarks \cite{orland-3+1}, but one of the couplings, namely $g_{0}^{\prime\prime}$, is 
large. This means that the $3+1$ dimensional theory can only be studied  by mixed 
weak/strong-coupling 
methods, unlike in $2+1$ dimensions, where no strong-coupling assumption is needed.

Some of the ideas discussed here were anticipated by others. The lattice formulation of
gauge theories as coupled sigma models was discussed more than three decades ago 
in the light-cone gauge (in contrast to our use of the axial gauge)
\cite{bardeen-pearson}. There were attempts by Griffin to use integrability in $1+1$ dimensions, in this context, to understand Yang-Mills fields \cite{griffin}. Durhuus and 
Fr\"{o}hlich estimated the potential energy of a quark and antiquark separated in the $2$-direction 
\cite{durhuus-fr} (though not the $1$-direction), which agrees with 
our leading result in powers of $g_{0}^{\prime}$ \cite{orland2+1}. See also Reference 
\cite{koma}. The respect in which our work is different is 
that we have successfully used this strategy to 
study physical quantities, namely 
the string tension and the mass spectrum, in detail.

Perturbing away from integrability to study field-theoretic and many-body systems is an active 
sub-field of
statistical and condensed-matter physics (for reviews, see \cite{bhaseen-tsvelik}). The first 
work in this field 
is that of McCoy and Wu for the Ising 
model \cite{mccoy-wu} and by Affleck and Weston \cite{affleck-weston} for spin chains.

\section{Gauge theories as coupled sigma models}

The formulation of the Yang-Mills theory as a collection of coupled sigma models has been discussed in detail elsewhere 
\cite{orland2+1,hor-string,mass-spectrum}, so we will only present the result here. For simplicity we discuss the $2+1$-dimensional Hamiltonian (in which $x^{2}$ is discrete) in this section, but 
the $3+1$-dimensional case \cite{orland-3+1} (for which $x^{1}$ and $x^{2}$ are discrete) is similar. We use axial gauge $A_{1}=0$, or $U_{1}=1$. The remaining lattice gauge field is $U_{2}(x^{0},x^{1}, x^{2})$, and we drop the subscript $2$. The coordinates $x^{0}$ and $x^{1}$ are continuous, but
$x^{2}$ is discrete (as mentioned earlier in this paragraph). The left-handed and right-handed currents are
$j^{\rm L}_{\mu}(x)_{b}={\rm i}{\rm Tr}\,t_{b} \, \partial_{\mu}U(x)\, U(x)^{\dagger}$ and
$j^{\rm R}_{\mu}(x)_{b}={\rm i}{\rm Tr}\,t_{b} \, U(x)^{\dagger}\partial_{\mu}U(x)$, respectively, 
where $\mu=0,1$. The Hamiltonian obtained from (\ref{cont-lag-2+1}) is $H_{0}+H_{1}$, where
\begin{eqnarray}
H_{0}\!=\!\sum_{x^{2}}\int dx^{1} \frac{1}{2g_{0}^{2}}\{ [j^{\rm L}_{0}(x)_{b}]^{2}+[j^{\rm L}_{1}(x)_{b}]^{2}\}
\;,\label{HNLSM}
\end{eqnarray}
and
\begin{eqnarray}
H_{1}\!\!&\!\!=\!\!&\!\! \sum_{x^{2}}  \int dx^{1} \,
\frac{(g_{0}^{\prime})^{2}a^{2}}{4}\,\partial_{1}\Phi(x^{1},x^{2})\partial_{1}\Phi(x^{1},x^{2}) \nonumber \\
\!\!&\!\!-\!\!&\!\! 
\left(\frac{g_{0}^{\prime}}{g_{0}}\right)^{2}\,\,\sum_{x^{2}=0}^{L^{2}-a}  \int dx^{1} \!\!
\left[ j^{\rm L}_{0}(x^{1},x^{2})\Phi(x^{1},x^{2}) -j^{\rm R}_{0}(x^{1},x^{2}) \Phi(x^{1},x^{2}+a) \right]  
\nonumber \\
&+&\;\;\;\;\;(g_{0}^{\prime})^{2}q_{b}\Phi(u^{1},u^{2})_{b} -(g_{0}^{\prime})^{2}
q^{\prime}_{b}\Phi(v^{1},v^{2})_{b}   \; ,
\label{continuum-local}
\end{eqnarray}
where $-\Phi_{b}=A_{0\,\,b}$ is the temporal gauge field, and
where in the last term
we have inserted two color charges - a quark with charge $q$ at site $u$
and an anti-quark with charge $q^{\prime}$ at site $v$. Some gauge invariance remains
after the axial-gauge fixing, namely that 
for each $x^{2}$
\begin{eqnarray}
\left\{ \int d x^{1}\left[ j^{L}_{0}(x^{1},x^{2})_{b}-j^{R}_{0}(x^{1},x^{2}-a)_{b}\right] - g_{0}^{2}Q(x^{2})_{b} \right\}\Psi=0\;,
\label{physical}
\end{eqnarray}
on wave functionals $\Psi$, 
where $Q(x^{2})_{b}$ is the total color charge from quarks at $x^{2}$ and $\Psi$ is any physical 
state. To derive the constraint (\ref{physical}) more precisely, we started with open boundary
conditions in the $1$-direction and periodic boundary conditions in
the $2$-direction, meaning that the two-dimensional space is a cylinder.

The unperturbed Hamiltonian (\ref{HNLSM}) is a discrete sum of principal-chiral nonlinear sigma model Hamiltonians with ${\rm SU}_{\rm L}(N)\times {\rm SU}_{\rm R}(N)$ symmetry. This sigma model is asymptotically free and has been argued to have a mass gap. Its basic excitations are soliton-like 
particles, labeled by index $r=1$, which can form bound states, labeled by an index 
$r=2,\dots, N-1$ (the $r=N-1$ excitation, a bound state of $N-1$ ``elementary" $r=1$ particles, is the antiparticle of the of the $r=1$ particles). These elementary excitations of the
sigma model are color dipoles, and can be thought of as bound pairs of 
chiral-Gross-Neveu Fermions. 

The S-matrix of two elementary excitations of the sigma model is \cite{abda-wieg}
\begin{eqnarray}
{\mathfrak S}_{11}(\theta)
=
\frac{\sin (\theta/2-\pi{\rm i}/N)}{\sin(\theta/2+\pi{\rm i}/N)}\;S_{\rm CGN}(\theta)\otimes 
S_{\rm CGN}(\theta) ,\label{s-matrix}
\end{eqnarray}
where $S_{\rm CGN}$ is the S-matrix of two elementary excitations of the chiral Gross-Neveu
model \cite{cgn}:
\begin{eqnarray}
S_{\rm CGN}(\theta)\!\!=\!\!\frac{\Gamma({\rm i}\theta/2\pi+1)\Gamma(-{\rm i}\theta/2\pi-1/N) }{\Gamma({\rm i}\theta/2\pi+1-1/N) \Gamma(-{\rm i}\theta/2\pi)}
\left( 1-\frac{2\pi{\rm i}}{N\theta}P\right), \label{cgn}
\end{eqnarray}
where $P$ switches the colors of the elementary Gross-Neveu particles. Other S-matrix elements can be found by crossing and by fusion techniques. The mass spectrum of the bound states of the elementary particles is 
$m_{r}=m_{1}\frac{\sin\pi r/N}{\sin \pi/N}$.

The physical interpretation is that transverse electric flux consists of the massive particles 
of the sigma model. These are joined by a lighter longitudinal electric flux (to satisfy Gauss' law) which is essentially that of the $1+1$ dimensional gauge theory. For finite $N$, the bound states are not free strings, but scatter nontrivially.

\section{Confinement in $2+1$-dimensions}

The exact S-matrix and current form factor were used to study the dependence of the mass spectrum and the quark-antiquark potential on $g_{0}^{\prime}$ for gauge group SU($2$).

From the current form factor, the string tension in the longitudinal \cite{hor-string} was found to be  
\begin{eqnarray}
\sigma_{\rm long}= 
\frac{3 (g_{0}^{\prime})^{4} }{8 K} \;, \label{hor-string-tension}
\end{eqnarray}
where the factor $K$ is given by
\begin{eqnarray}
K=\frac{(g_{0}^{\prime})^{2}a^{2}}{4}+
\frac{1}{3m^{2}\pi^{2}}\left(\frac{g_{0}^{\prime}}{g_{0}}\right)^{4}
\exp\left[-2\int_{0}^{\infty} \frac{d\xi}{\xi} {e^{-\xi}}{\tanh^{2}\frac{\xi}{2}}\right]  \;,
\label{constant-of-field-squared}
\end{eqnarray}
$m$ is the mass gap and $a$ is the lattice spacing. The string tension in the transverse direction \cite{vert-string} is 
\begin{eqnarray}
\sigma_{\rm trans}=\frac{m}{a}-\frac{2{\sqrt 3}}{\pi}\frac{g_{0}^{\prime}}{g_{0}^{2}a^{2}}\;, 
\label{vert-string-tension}
\nonumber
\end{eqnarray}
where $m$ is the sigma-model mass gap,
respectively. The  terms of order $(g_{0}^{\prime})^{4}$
in (\ref{hor-string-tension}) come from transverse oscillations of the string. The second term in
(\ref{vert-string-tension}) comes from the smearing of color of each transverse string constituent (that is sigma-model excitation) over a region of size $m^{-1}$.

It is possible to find the spectrum of glueball states \cite{mass-spectrum}. The bound states consist of an elementary sigma-model particle and antipartcle bound by two lines of longitudinal electric flux (it has the topology of a ring). The role of the S-matrix is to determine the
matching condition of the bound-state wave 
function at the origin. The spectrum 
can be worked out using the WKB method. The result is 
\begin{eqnarray}
M_{n}=2m+E_{n}=2m+\left[
\epsilon_{n}^{1/3}- \frac{3(3-2\ln2)\sigma_{\rm long} }{ 4\pi m} \epsilon_{n}^{-1/3}
\right]^{2},\;\;n=0,1,2,\dots \label{bound-state-spectrum}
\end{eqnarray}
where
\begin{eqnarray}
\epsilon_{n}=\frac{3\pi\sigma_{\rm long}( n+\frac{1}{2} ) }{4m^{1/2}}
+\left\{
\left[
\frac{3\pi\sigma_{\rm long}}{4m^{1/2}(n+\frac{1}{2})}
\right]^{2}
+\frac{1}{8}
\left[
\frac{3(3-2\ln2)\sigma_{\rm long}}{2\pi m}
\right]^{3}
\right\}^{1/2}.
\label{epsilon-definition}
\end{eqnarray}
This was found for the SU($2$) gauge group, but it is elementary to generalize the result to any 
SU($N$).

What is interesting about these results is that they show there is no deconfining phase transition as $g_{0}^{\prime}$ is increased from zero. The leading terms in $g_{0}^{\prime}$ in all these expressions are just those found from elementary arguments \cite{orland2+1,durhuus-fr}. There is, however, a dimensional cross-over \cite{hor-string}. The 
cross-over is a change from $1+1$-dimensional behavior
to $2+1$-dimensional behavior as $g_{0}^{\prime}$ is increased. This problem is still under investigation. In the case of the ${\mathbb Z}_{2}$ gauge theory, dual to the three-dimensional Ising model, the same type of cross-over has been overcome using the density-matrix renormalization group 
\cite{konik} (in fact, the formulation of the three-dimensional Ising model as coupled two-dimensional Ising models is very similar to the formulation of the $2+1$-dimensional gauge theory
as coupled $1+1$-dimensional sigma models discussed here). Realistic results for the 
correlation-length critical exponent $\nu$ were obtained this way.

\section{Longitudinal rescaling in $3+1$ dimensions}

As it happens, the naive classical rescaling of coordinates $x^{0,3}\rightarrow \lambda x^{0,3}$, $x^{1,2}\rightarrow x^{1,2}$, mentioned 
in the introduction, is not how quantum field theories actually rescale. There are anomalous dimensions, just as there are for dilatations. 

There is a straightforward Wilsonian procedure to determine how the Lagrangian changes under a longitudinal 
rescaling. Suppose that there is a spherical momentum cut-off $\Lambda$ on the Fourier components of the gauge field (counterterms are 
needed to restore gauge invariance). Let $\tilde \Lambda$ be less than or equal to
$\Lambda$. Consider the ``fast" 
degrees of freedom, whose momenta outside of a four-dimensional ellipsoid with two major axes of
$2{\tilde \Lambda}$ and two minor axes of
$2{\tilde \Lambda}\lambda$, but inside the four-dimensional sphere of diameter $2\Lambda$. These fast degrees of freedom are integrated out
of the functional integral. Finally a longitudinal rescaling by $\lambda$ restores a spherical
cut-off, but now of diameter $2\tilde \Lambda$. The one-loop result is \cite{orland-xiao}
\begin{eqnarray}
{\mathcal L}_{\rm eff}=\frac{1}{ 4g_{\rm eff}^{2} }
\,{\rm Tr}\,\left(
{F}_{01}^{2}+{F}_{02}^{2}-{F}_{13}^{2}-{F}_{23}^{2}
+\lambda^{-2+\frac{17C_{N}}{48\pi^{2}}{\tilde g}_{0}^{2}}{F}_{03}^{2}-
\lambda^{2+\frac{7C_{N}}{48\pi^{2}}{\tilde g}_{0}^{2}}{F}_{12}^{2}
\right)+\cdots
\;,\label{effective-lag}
\end{eqnarray}
where 
\begin{eqnarray}
\frac{1}{g_{\rm eff}^{2}}=\frac{1}{g_{0}^{2}} 
-\frac{11C_{N}}{48\pi^{2}} \ln \frac{\Lambda}{\tilde \Lambda}
+\frac{C_{N}\ln {\lambda}}{32\pi^{2}}
\;. \label{effective-coupling}
\end{eqnarray}
The corrections to (\ref{effective-lag}) are of order $\ln \lambda^{2}$. 

What (\ref{effective-lag}) shows is that one cannot really trust perturbation theory to find the
effective longitudinally-rescaled action at {\em very} high energies. The point is that 
(\ref{effective-lag}) is really only valid at weak coupling, as the corrections from the anomalous dimension are not significant. Large rescalings generate a 
large coupling associated with ${\rm Tr} F_{12}^{2}$. The coefficient of this term is small, meaning that the longitudinal magnetic field is wildly fluctuating. We cannot take this action too seriously for small $\lambda$, just as we cannot take strong-coupling approximations in
lattice or AdS-type QCD theories too seriously. A ``good" strongly-coupling theory is one in which 
we have somehow correctly integrated out the high-momentum degrees of freedom over very large ranges of momentum. For further discussion of this point, see Reference \cite{orland-xiao}.

Since we cannot rigorously extend the expression (\ref{effective-lag}) to $\lambda \ll1 $, the best 
we can do is guess the form of an effective theory at high energy. Probably the most sensible approach is to replace the sigma model of Reference \cite{orland-3+1} by an arbitrary $1+1$-dimensional interacting massive
field theory with global ${\rm SU}(N)_{\rm L}\times {\rm SU}(N)_{\rm R}$ symmetry, coupling its currents together to $A_{0}$. Though not QCD, this theory 
has local SU($N$) gauge symmetry. Such an effective 
theory certainly exhibits a forward peak, and may be a good model of soft scattering. 

\section{Some new directions}

In $2+1$ dimensions, the main problem is to overcome the cross-over, to understand the isotropic
case. As mentioned already, this has been accomplished for the ${\mathbb Z}_{2}$ case
\cite{konik}. The problem should perhaps be easier for SU($N$) theories, as the critical point
is the same for both the isotropic and anisotropic theories; it is simply at 
$g_{0}=g_{0}^{\prime}=0$. 

Though string tensions have been studied for arbitrary SU($N$)
\cite{arbN} and the mass spectrum is possible to determine (though this has not been published yet) the corrections to the string tensions in powers of $g_{0}^{\prime}$ cannot be found until the form factors of the sigma model are determined. We have made some progress on the $1/N$-expansion of these form factors (the bound-state structure is much simpler for large $N$). This should make it possible to study string dynamics more generally, as the sigma S-matrix becomes trivial in this limit. 

It seems worthwhile to study effective SU($N$) gauge theories discussed at the end of the last section in $3+1$ dimensions. These are effective parton models of soft scattering. In particular, it appears that the AGK cutting rules are valid 
\cite{AGK}. Whether
other useful results can be obtained is an open question.

\end{document}